\documentclass [10pt,twoside]{article}

\usepackage{shrthnds}
\usepackage{cite}
\usepackage{graphicx}
\usepackage[usenames]{color}
\usepackage{subfigure}
\usepackage{setspace}

\usepackage{multirow}

\usepackage[hang,small]{caption}

\usepackage{amsmath}                  %text inside eqnarray

\usepackage{geometry}
\usepackage{fancyhdr}
\usepackage{titlesec,titletoc}
  \titleformat{\section}{\Large\sf\bfseries}{\thesection}{1em}{}
  \titleformat{\subsection}{\large\sf\bfseries}{\thesubsection}{1em}{}

\title{\sf\bfseries \ntitle}
\author{ Pankaj Jain$^1$\footnote{email: pkjain@iitk.ac.in}~, 
 Atul Jaiswal$^1$\footnote{email: atijazz@iitk.ac.in}~, 
Purnendu Karmakar$^1$\footnote{email: purnendu@iitk.ac.in}~,\\
Gopal Kashyap$^1$\footnote{email: gopal@iitk.ac.in}~
and Naveen K. Singh$^2$\footnote{email:  naveenks@prl.res.in}}
\date{}%{\today}

\newcommand{\pghdr}{\footnotesize {P. Jain} {\it et al.} -- Cosmological Implications of Unimodular Gravity}
\newcommand{\ntitle}{Cosmological Implications of Unimodular Gravity}

\begin{document}
\vspace{-3cm}
\maketitle
%\vspace{-1.5cm}
\vspace{-0.6cm}
\bc
\small{{}$^1$Department of Physics, IIT Kanpur, Kanpur 208 016, India\\
{}$^2$ Physical Research Laboratory, Ahmedabad 380 009, India}
%{\small 1) Aff 1\\
%2) Aff 2\\
%3) Aff 3}
\ec

\bc\begin{minipage}{0.9\textwidth}\begin{spacing}{1}{\small {\bf Abstract:}
%%%%%%%%%%%%%%%%%%%%%%%%%%%%%%%%%%%%%%%%%%%%%%%%%%%%%%
% ABSTRACT
%%%%%%%%%%%%%%%%%%%%%%%%%%%%%%%%%%%%%%%%%%%%%%%%%%%%%%
We consider a model of gravity and matter fields which is invariant
only under unimodular general 
coordinate transformations (GCT). The determinant of the metric is treated
as a separate field which transforms as a scalar under unimodular GCT. 
Furthermore we also demand that the theory is invariant under a new
global symmetry which we call generalized conformal invariance.
We study the cosmological implications of the resulting theory. 
We show that this theory gives a 
 fit to the high-z supernova data which is identical to the standard Big
Bang model. Hence we require some other cosmological observations to 
test the validity of this model. We also consider some models which do not
obey the generalized conformal invariance. In these models we can fit the
supernova data without introducing the standard cosmological constant term.
Furthermore these models introduce only one dark component and hence
solve the coincidence problem of dark matter and dark energy.

%%%%%%%%%%%%%%%%%%%%%%%%%%%%%%%%%%%%%%%%%%%%%%%%%%%%%%
}\end{spacing}\end{minipage}\ec

%\vspace{0.5cm}\begin{spacing}{1.1}

\section{Introduction}
In a recent paper \cite{Jain:2011} we have considered some implications of 
a model which is invariant
only under the restricted unimodular general coordinate transformations (GCT)
but not the full GCT.
The basic idea that the gravitational action may be invariant only 
under the unimodular GCT 
 was first proposed by 
Anderson and Finkelstein
\cite{Anderson}. It is related to an earlier proposal by Einstein
\cite{Einstein}. In this proposal the determinant of the metric, $g$, is not a
dynamical variable. 
This idea has been pursued in detail in many papers
\cite{Ng1,Zee,Buchmuller,Henneaux89,Unruh1989,Ng2,Finkelstein,Alvarez05,Alvarez06,Ellis},
which have considered its application to the problem of the
cosmological constant and its quantization.
Since $g$ is not a dynamical field, a model based on unimodular GCT 
has the potential to solve the cosmological 
constant problem. However  
as explained in \cite{Weinberg}, 
the problem is 
not really solved. 

In the present paper we treat the determinant as an independent 
scalar field since we only demand
invariance under unimodular general coordinate transformations (GCT)
 \cite{Buchmuller,Weinberg,Alvarez07,JMS,Alvarez09,Alvarez10,Shaposhnikov}.  
We study a class of such models. We first study a model 
  is invariant under the global conformal 
transformations \cite{Birrell}. This restricts the action considerably.  
We show that the resulting model is ruled out since it predicts null
redshift. We next define a class of models which are invariant under
a new global symmetry which we call generalized conformal invariance. 
We obtain the redshift dependence of luminosity distance in these models
by including dark matter and vacuum energy.  
We also consider two additional models based on unimodular gravity
which fit the high $z$ supernova data in terms of a single dark component. 
Hence these models solve the cosmic coincidence problem of dark matter
and dark energy. Furthermore these models do not require the standard
cosmological constant term and hence alleviate the fine tuning problem
of the cosmological constant.

We also need to verify that the models we study admit 
the standard spherically symmetric Schwarzchild solution so as to agree with data on solar system 
scale. In an earlier paper \cite{Jain:2011}
we have shown that this is true explicitly 
for the case of a particular model based on unimodular gravity.
In general the solution is expected to differ in different models. 
However the difference
arises only due to the field $\chi$ which represents cosmic evolution. 
Such a field can show deviations only over very large distances and hence
cannot significantly affect the physics on the scale of the solar system.
We demonstrate this by obtaining an explicit solution for one of the 
models considered in this paper. 
Such a solution is also studied in a particular class of unimodular models 
in Ref. \cite{SinghNK}.

Several of the models that we study in this paper effectively involve mass
parameters which evolve on cosmological time scales. Hence the masses of
all particles, such as electrons, protons, as well other mass parameters,
such as the gravitational constant and Fermi constant evolve slowly with
time. Hence the derivation of the standard Hubble law as well as the
luminosity distance is somewhat more involved in these models in comparison
to the standard Big Bang model. Furthermore here we have to be sure that
we do not disagree with some other observables such as the supernova stretch
factors \cite{Goldhaber}. We address all these issues in this paper.

\section{Review of Unimodular Gravity}
We require only unimodular general
coordinate transformations (GCT), which are defined by,
\be
x^\mu \rightarrow x'^{\mu}
\ee  
such that 
\be
{\rm det}(\partial x'^\mu/\partial x^\nu)=1  
\ee
It is convenient to split 
the standard metric as follows
\cite{Jain:2011},
\begin{eqnarray}
g_{\mu\nu}&=&\chi^2 \bar{g}_{\mu\nu}
\label{eq:gmunu}
\end{eqnarray}
where the determinant $\bar g$ of $\bar{g}_{\mu\nu}$ is assumed to be 
non-dynamical. Hence we demand that $\bar g$ is fixed such that, 
\begin{eqnarray}
\bar g = det[\bar {g}_{\mu\nu}]= f(x) 
\label{eq:constraint}
\end{eqnarray}
where $f(x)$ is some function of the space-time coordinates. 
The field $\chi$ behaves as a scalar field under unimodular GCT. 
Hence the basic fields of our theory are $\chi$, the metric $\bar g_{\mu\nu}$
and matter fields. We denote the connection, the Ricci tensor and the
curvature scalar by the symbols $\bar \Gamma^\mu_{\alpha\beta}$, 
$\bar R_{\mu\nu}$ and $\bar R$ respectively. All these quantities are
computed by using the metric $\bar g_{\mu\nu}$.

\section{Unimodular gravity with global conformal invariance}
We next present a model of gravity and matter fields which is invariant 
under unimodular GCT but not the full GCT. As discussed in 
\cite{Jain:2011,Buchmuller,Weinberg,Alvarez07,Shaposhnikov}
there is considerable freedom in writing such a model. We impose a further
constraint on this model that it should satisfy global conformal invariance
\cite{Birrell}.  
Under this transformation, the
coordinates $x^\mu$ do not change. However the metric and the fields undergo
suitable transformation. In 4 space-time dimensions, 
the full metric essentially transforms as
\be
g_{\mu\nu} \rar g_{\mu\nu}\Omega^2
\ee 
where $\Omega$ is a constant parameter. This transformation changes the
determinant of the metric. Hence in our case we may express this
transformation as,
\ba
\bar g_{\mu\nu} &\rar & \bar g_{\mu\nu}\nonumber\cr
\chi &\rar & \chi \Omega
\ea
The matter fields transform as follows,
\ba
\phi &\rar & \phi/\Omega\nonumber\cr
A_\mu &\rar & A_\mu  \nonumber\cr
\psi &\rar & \psi/\Omega^{3/2}
\ea
where $\phi$, $A_\mu$ and $\psi$ are scalar, vector and spinor fields 
respectively.

The action invariant under unimodular GCT and global conformal transformations
may be written as,
\be
S = \int d^4x \sqrt{-\bar g}\left[{1\over \kappa} \bar R - {\xi\over \kappa}
\bar g^{\mu\nu} \partial_\mu\ln\chi\, \partial_\nu\ln\chi\right] +S_M 
\label{eq:actionS}
\ee
where $S_M$ represents the matter part of the action and 
$\kappa = 16\pi G$. 
It is useful to compare this action with the standard gravitational action
written in terms of $\bar g_{\mu\nu}$ and $\chi$ \cite{Jain:2011}. 
We see that the current action represents a significant modification
of the standard Einstein's action. Hence it is likely to give very
different predictions and should be carefully examined to see if it agrees
with observations. As we show below, this model does not
lead to any redshift. This prediction is rather counter intuitive since the
model does lead to expansion, as we shall see in section 4. 
However null redshift implies that the model
is ruled out.  

The analysis of redshift in this model is a little more complicated in 
comparison to that in the standard Einstein's gravity. As we shall see,
it is not reasonable to assume that the emitted atomic frequencies in the early 
Universe are identical to those emitted by atoms today. We first write 
down the matter action in terms of the fields $\bar g_{\mu\nu}$ and
$\chi$.  
 Before doing that we also need to suitably split
the veirbien field to be consistent with Eq. \ref{eq:gmunu}. 
We consider the veirbien field $e_i^a$ where $a$ represent the Lorentz index.
The full metric
\begin{equation}
g_{\mu\nu} = e^a_\mu \eta_{ab} e^b_\nu
\end{equation}
We may split
\begin{equation}
e^a_\mu = \chi \bar e^a_\mu
\end{equation}
Here we have defined $\bar e^a_\mu$ such that it has determinant equal
to $\sqrt{-{\bar g}}$.
We now write down the action for a real scalar field, $\phi$, a dirac fermion, 
$\psi$, and the electromagnetic field, $A_\mu$. 
Generalization to the more realistic case
of the Standard model is straightforward but not necessary for our purpose.
The action may be written as,
\begin{equation}
\mathcal{S}_M = \int d^4x \sqrt{-\bar g}\Bigg[ {\chi^2\over 2}
\bar g^{\mu\nu} (\partial_\mu \phi)(\partial_\nu \phi) - 
{\chi^2\over 2} m^2_1\phi^2 -\frac14
\bar g^{\mu\nu}\bar g^{\alpha\beta}({F}_{\mu\alpha} {F}_{\nu\beta})\Bigg]
 + {\cal S}_{\rm fermions},
\label{eq:S_matter}
\end{equation}
where $F_{\mu\nu}$
is the electromagnetic field strength tensor and $m_1$ the scalar particle 
mass.
The fermion action may 
be expressed as,
\begin{equation}
{\cal S}_{\rm fermions} = \int d^4 x\, \bar e\, \left(\chi^3{\overline\psi}i
\gamma^\mu  {\cal D_\mu} \psi - \chi^3 m_2 \bar\psi\psi 
 \right)
\label{eq:S_fermions}
\end{equation}
where $e=\sqrt{-\bar g}$, 
$\gamma^\mu=\bar e^\mu_{~a}\gamma^a$  and
$a, b$ are Lorentz indices.
Here 
$m_2$ is the fermion mass. 
The covariant derivative acting on the fermion field is defined by
\begin{equation}
{\cal D}_\mu \psi = \left({\tilde D}_\mu + {1\over2}\omega_{\mu}^{ab}\sigma_{ab}\right)\psi\ ,
\label{eq:Dfermion}
\end{equation}
where ${\tilde D}_\mu$ is the electromagnetic gauge covariant derivative,
$\sigma_{ab} = {1\over4}[\gamma_a,\gamma_b]$ and $\omega_{\mu}^{ab}$ is the spin
connection. 
We point out that the matter action is essentially the same as in the case
of generally covariant model, except for a crucial difference.
The mass terms contain a different power of $\chi$. For the case of general
covariance, both the fermion and scalar mass terms would have a factor of
$\chi^4$. 
 These mass terms breaks GCT but preserve global conformal invariance. 

We next obtain the free electromagnetic wave solutions in this theory.
Here we should interpret the time coordinate as the conformal time. 
Hence we denote it with the symbol $\eta$. 
The electromagnetic action in our theory is the same as in generally covariant
theory. In the present case we shall
take $\bar g_{\mu\nu} = \eta_{\mu\nu}$, the Lorentz metric. Hence in these
coordinates the electromagnetic action is the same as that in flat space-time.
The free electromagnetic wave solutions in this case may be written as
\begin{equation}
A_\mu(\eta) \propto A_\mu(0) e^{-i\omega_e \eta} = A_\mu(0) e^{-i\omega_e 
\int{dt/ \chi(t)}}
\label{eq:EMWave}
\end{equation}
where $t$ is the cosmic time and $\omega_e$ is a constant related
to the emitted frequency. The frequency at any time may be obtained
by taking the time derivative of the exponent. We obtain
\begin{equation}
\omega(t) = {\omega_e\over \chi(t)}
\label{eq:omegat}
\end{equation}
This frequency at current time needs to be compared with the frequency 
of the waves emitted by current atomic transitions. 

In order to proceed 
further we need to determine $\omega_e$.  
This is facilitated by obtaining an effective action of fermions coupled to
the electromagnetic field. We may define a scaled field $\psi' =
\chi^{3/2}\psi$. In terms of the scaled field, we obtain,
\begin{equation}
{\cal S}_{\rm effective} = \int d^4 x\, \bar e\, \left({\overline{\psi'}}i
\gamma^\mu  {\cal D_\mu} \psi' -  \bar m_2 {\overline{\psi'}}\psi' 
 \right) + ...
\label{eq:effectiveA}
\end{equation}
where we have not explicitly displayed the term proportional to the time
derivative of $\chi$. This term is suppressed by $H/m$ in comparison 
to the terms we keep, where $H$ is the Hubble constant. 
The term $\bar m_2$ in the present case is simply equal to $m_2$. 
For models considered later, 
$\bar m_2$ will be equal to $m_2$ times some power of the scale factor $\chi$. 
Hence we obtain
an effective action which is simply the same as the action in flat space-time.
The time coordinate in this action is of course just the conformal time. 
None of the parameters of this effective action scale with $\chi(t)$. 
Hence the electromagnetic wave solution to this is given by Eq. \ref{eq:EMWave}
with the frequency $\omega_e$ independent of the time when the 
wave is emitted. The observed frequency at any time after emission is 
given by Eq. \ref{eq:omegat}.  This frequency observed today has to be 
compared with the frequency of atomic transitions at current time. That 
would be given by $\omega'(t) = \omega_e/\chi(t)$. We find that, since
$\omega_e$ is independent of the time of emission, $\omega'(t) = \omega(t)$. 
The standard definition of redshift, $z$, is
\begin{equation}
{\omega'(t_0)\over \omega(t_0)} = 1+z
\end{equation} 
where $t_0$ is the current time. Hence we find that the 
conformal invariant
unimodular model gives null redshift. The model is, therefore, ruled out
by cosmological data.

Our analysis of the Hubble law is slightly different from the standard
textbook derivation. Hence it is useful to verify that it gives the 
expected answer
for the standard case of Einstein's gravity. The only difference in the
matter action in this
case is that the mass terms in Eqs. \ref{eq:S_matter} and 
Eq. \ref{eq:S_fermions} have a factor of $\chi^4$. Hence
in the effective action, Eq. \ref{eq:effectiveA}, 
the scaled mass
$\bar m_2 = m_2\chi$. This implies that the factor $\omega_e$ in  
Eq. \ref{eq:EMWave} should be scaled by the factor $\chi(t_e)$, where
$t_e$ is the time of emission. Hence in the present case 
the observed frequency due to a transition at early time is 
$\omega(t_0) = k\chi(t_e)/\chi(t_0)$, whereas the frequency due a transition
in laboratory is $\omega'(t_0) = k$, where $k$ is a constant. Hence, as
expected, $\omega'(t_0)/\omega(t_0) = \chi(t_0)/\chi(t_e)=1+z$. This will,
of course, predict non-zero redshift.

\section{Generalized Conformal Invariance}
Our conformal invariant model, discussed in section 3, 
might have been interesting since
it predicts exactly zero cosmological constant. However it fails in a 
much more dramatic way since it leads to null redshift. In the present
section we determine if we can modify this theory such that this null
result may be avoided. We shall show that it is possible to define a 
range of models, labelled by a continuous parameter, which are invariant
under a generalized conformal transformation.

Let us consider the following modified action, 
\be
S = \int d^4x \sqrt{-\bar g}\left[{\chi^{2\alpha}\over \kappa} \bar R - 
{\xi\over \kappa}
\bar g^{\mu\nu}\chi^{2\zeta} \partial_\mu\chi\, \partial_\nu\chi\right] +S_M 
+S_\Lambda
\label{eq:actionGS}
\ee
with 
\begin{equation}
\mathcal{S}_M = \int d^4x \sqrt{-\bar g}\Bigg[ {\chi^2\over 2}
\bar g^{\mu\nu} (\partial_\mu \phi)(\partial_\nu \phi) - 
{\chi^{4\beta}\over 2} m_1^2\phi^2 -\frac14
\bar g^{\mu\nu}\bar g^{\alpha\beta}({F}_{\mu\alpha} {F}_{\nu\beta})\Bigg]
 + {\cal S}_{\rm fermions},
\label{eq:GS_matter}
\end{equation}
\begin{equation}
{\cal S}_{\rm fermions} = \int d^4 x\, \bar e\, \left(\chi^3{\overline\psi}i
\gamma^\mu  {\cal D_\mu} \psi - \chi^{3\gamma} m_2 \bar\psi\psi 
 \right)
\label{eq:GS_fermions}
\end{equation}
and 
\begin{equation}
S_\Lambda = -\int d^4x\sqrt{-\bar g} \Lambda \chi^\delta
\label{eq:Lambda}
\end{equation}
where $\Lambda$ is the cosmological constant and the exponents $\alpha$, $\beta$, $\gamma$, $\delta$ and $\zeta$ are constant parameters. 
We note that for the case of general relativity the parameters,
$\alpha=1$, $\beta=1$, $\delta=4$, $\gamma=4/3$, $\zeta=0$ and $\xi=6$. We shall allow these 
parameters to deviate from these values such that the model still
satisfies a generalized global conformal invariance. We expect that for
consistent cosmological evolution the parameter $\xi$ will differ from 
6 and get fixed in terms of the other parameters. 
We point out that the kinetic terms of the scalar, fermion and vector
fields are exactly the same as in general relativity. 
The power of
$\chi$ in the scalar field kinetic energy term can be fixed by a suitable
definition of $\chi$. Once we have fixed the scalar kinetic energy term 
it is reasonable to choose the power of $\chi$ in the fermion kinetic
energy term as given in Eq. \ref{eq:GS_fermions}. If we choose a different
power then in the non-relativistic limit the cosmological
implications of fermions will be different from those of bosons. 
This is because of the constraint imposed by generalized
conformal invariance. At present
we do not allow this possibility. Finally the vector field kinetic energy
term does not depend on $\chi$. If this term also depends on $\chi$, then,
as we shall see,
it will lead to cosmic evolution of the electromagnetic gauge coupling.

We fix the different exponents of $\chi$ by demanding that the action
is invariant under the transformation,
\begin{equation}
\chi\rar \chi/\Omega\ , \ \ \ \ x\rar \Omega^a x\ ,\ \ \ \ \ \phi\rar\Omega^b
\phi\ ,\ \ \ \ \ \psi\rar\Omega^c\psi\ ,\ \ \ \ \ A_\mu\rar A_\mu/\Omega^a
\end{equation}
where $\Omega$ is a constant. We find that the invariance of the term 
proportional to $\bar R$ in Eq. \ref{eq:actionGS} requires that $a=\alpha$.  
Invariance of the scalar and fermion field terms in the action 
require that 
\begin{equation}
b=1-\alpha\ ,\ \ \ \  \beta=(1+\alpha)/2\ ,\ \ \ \  
c= 3(1-\alpha)/2\ ,\ \ \ \ \gamma = 1+(\alpha/3) 
\end{equation}
We next point
out that the vector field scales exactly as the partial derivative
$\partial_\mu$ under the generalized conformal transformation. This is true
as long as the vector field kinetic term is independent of $\chi$. 
Alternatively if this term depended on $\chi$, the transformation of
$A_\mu$ would be different and the gauge covariant derivative would 
depend on $\chi$, effectively leading to a cosmic evolution of the gauge
coupling. In the present paper we do not allow this possibility. 
Finally the invariance of $S_\Lambda$ and the kinetic energy term for
$\chi$ requires that 
\begin{equation}
\delta = 4\alpha \ , \ \ \ \ \ \zeta = \alpha-1
\end{equation}

We next study the cosmological implications of our unimodular theory with
generalized conformal invariance. We first obtain the formula for the 
cosmological redshift, which, as we shall show, deviates from the standard
result. We next derive the equations governing cosmological evolution
and apply these to compute the time dependence of the scale factor. 
Finally we fit the high redshift 
supernova Type 1a data \cite{Amanullah} to determine the parameters
of this model. 

\subsection{Cosmological Redshift}
We next obtain the relationship between the scale factor and the cosmological
redshift in our generalized model. The effective action for this model is
given by Eq. \ref{eq:effectiveA} with $\bar m_2 =\chi^\alpha m_2 $. 
In this case the current value of the frequency of a wave emitted at early 
a time $t_e$ would be 
\begin{equation}
\omega(t_0) = {k\chi^\alpha(t_e)\over \chi(t_0)} 
\end{equation}
where $k$ is a constant. This has to be compared with 
the frequency of a wave originating in the laboratory, given by,
\begin{equation}
\omega'(t_0) = {k\chi^\alpha(t_0)\over \chi(t_0)} 
\end{equation}
Hence we find that
\begin{equation}
{\omega(t_0)\over \omega'(t_0)} = \left({\chi(t_e)\over \chi(t_0)}\right)^\alpha
= {1\over 1+z} 
\label{eq:redshift}
\end{equation}
This deviates from the standard formula in Einstein's gravity by the 
extra power of $\alpha$ in the ratio of the scale factors.

There is another instructive way to verify the redshift formula, Eq. \ref{eq:redshift}. We define an effective action by scaling the mass parameter such
that the action formally appears the same as the action in the limit
of full covariance. Hence we now scale the mass term as
\be
\chi^{2+2\alpha} m_1^2 \phi^2 \rightarrow \chi^4 \tilde m_1^2\phi^2
\ee
where we have set $4\beta = 2+2\alpha$. On the right hand side we have 
the standard covariant mass term with a time varying mass
$\tilde m_1 = \chi^{\alpha -1} m_1$. Due to this time variation in mass,
the frequency of atomic transitions changes in proportion to the mass. Let
$\omega'(t_0)$ be the frequency of a laboratory transition corresponding
to the frequency $\omega(t_e)$ in the early Universe. The ratio,
\be
{\omega(t_e) \over \omega'(t_0)} = {\tilde m(t_e)\over \tilde m(t_0)}
= \left({\chi(t_e)\over \chi(t_0)}\right)^{\alpha-1}
\ee 
Hence we find
\be
{\omega(t_0)\over \omega'(t_0)} = {\omega(t_0)\over \omega(t_e)}
{\omega(t_e)\over \omega'(t_0)} = {\chi(t_e)\over \chi(t_0)}
\left({\chi(t_e)\over \chi(t_0)}\right)^{\alpha-1} 
\ee
which again leads to Eq. \ref{eq:redshift}.

\subsection{Cosmological Evolution}
The generalization of Einstein's equations to the
present theory, is given by, 
\ba
-\chi^{2\alpha}\left[\bar R_{\mu\nu} - 
{1\over 4} \bar g_{\mu\nu}\bar R\right] -
\left[\left(\chi^{2\alpha}\right)_{;\mu;\nu} - 
{1\over 4} \bar g_{\mu\nu} \left(\chi^{2\alpha}\right)_{;\lambda}^{;\lambda}\right] 
+\xi\chi^{2\zeta}\left[\partial_\mu\chi\, \partial_\nu\chi -
{1\over 4}\bar g_{\mu\nu}\,
\partial^\lambda\chi\, \partial_\lambda\chi \right] \nonumber\\
 = {\kappa\over 2} \left[T_{\mu\nu}
- {1\over 4} \bar g_{\mu\nu} T^\lambda_\lambda\right] 
\label{eq:Rmunu}
\ea
Here $T_{\mu\nu}$ represents all the contributions to this equation
obtained by the matter action. We may call this the energy momentum
tensor. However we caution the reader that it does not satisfy 
the usual conservation law.
The equation of motion for $\chi$ may be written as
\be
2\alpha\chi^{2\alpha-1}\bar R +  2\xi\zeta\chi^{2\zeta-1} 
\bar g^{\mu\nu}\partial_\mu\chi\partial_\nu\chi + 
2\xi \chi^{2\zeta} \bar g^{\mu\nu}(\chi)_{;\mu;\nu} = \kappa T_\chi
\label{eq:chi}
\ee
where $T_\chi$ represents the contributions to this equation due to the
matter fields. 

We shall assume, for simplicity, that 
\be
\bar g_{\mu\nu} = {\rm diagonal}[1,-1,-1,-1]
\ee
Hence the entire dynamics is contained in the field $\chi$. Here we shall
now determine the cosmological evolution assuming that 
the energy density of the universe is dominated by a particular component, 
radiation or non-relativistic matter.
Later we shall fit the high $z$ supernova data assuming non-relativistic
matter and vacuum energy.
 
\subsection{Radiation dominated Universe}
In the case of radiation dominated Universe, we find that,
\be
T_\chi = 0
\ee
We see this easily from the matter action. For the case of a vector field, 
there is no direct coupling to $\chi$. Hence its contribution to $T_\chi$
vanishes trivially. In the case of scalar or spinor field, it vanishes once
we impose the condition $P^2=0$. Hence in this case Eq. \ref{eq:chi} gives,
\be
{d^2\over d\eta^2} \chi = {(1-\alpha) \over \chi}\left({d\chi\over d\eta}\right)^2
\label{eq:ddlnchi}
\ee  
Here $\eta$ is the time coordinate. Note that the time coordinate in our
metric essentially corresponds to the conformal time in the standard Big Bang
cosmology. 
This implies that,
\begin{equation}
{d\chi\over d\eta} = C\chi^{1-\alpha}
\label{eq:rad1}
\end{equation}
where $C$ is a constant. 
We may express evolution of $\chi$ in terms of cosmic time, t, by using,
\be
\chi d\eta = dt
\ee
In terms of cosmic time we find that
\be
\chi\propto t^{1/(1+\alpha)}
\ee

We next determine the evolution of the energy density, $\rho$,
 by using Eq. \ref{eq:Rmunu}. 
In this equation $\bar R_{\mu\nu}=0$, $\bar R=0$. We may identify the energy momentum
tensor by taking the specific example of a scalar field \cite{Jain:2011},
\be
T_{\mu\nu} = \chi^2 \left<\partial_\mu\phi \partial_\nu\phi\right>
\label{eq:Tmunu}
\ee 
where the expectation value is taken in an appropriate thermal state
corresponding to the temperature of the medium. Furthermore, as discussed
in Ref. \cite{Jain:2011} 
\be
 \left<\partial_0\phi\partial_0\phi\right> = \chi^2(\eta)\rho
\label{eq:ExpKE}
\ee 
where $\rho$ is the energy density of the radiation field. We also find
that the trace, $T_\lambda^\lambda=0$.  
Substituting in Eq. \ref{eq:Rmunu}, we find 
\be
\rho = {8\alpha^2\over 3\kappa\chi^4}
\label{eq:rho}
\ee
Hence we find the standard result, $\rho\propto 1/\chi^4$. However
the time dependence of the scale factor is different for $\alpha\ne
1$.

\subsection{Non-relativistic Matter dominated Universe}
We next determine the cosmic evolution assuming that the energy
density is dominated by non-relativistic matter. 
Here we again consider a scalar field in the non-relativistic limit.
Other fields are expected to give the same result.

The factor $T_\chi$ in Eq. \ref{eq:chi} in this case is given by,
\be
-T_\chi = \chi \left<\bar g^{\mu\nu}\partial_\mu\phi\partial_\nu\phi\right>
-2\beta\chi^{4\beta-1} \left<m_1^2\phi^2\right>
\ee
Here only the time derivatives of $\phi$ are non-zero. 
The expectation value of the mass term is given by,
\be
\left<\chi^{4\beta}m_1^2\phi^2\right> = \chi^4\rho \ .
\label{eq:ExpMass}
\ee
We may obtain this by determining the effective Hamiltonian density
$\cal H$ in Minkowski space, as discussed in Ref. \cite{Jain:2011}. 
We scale the
field $\phi$ and the mass $m_1$ with an appropriate power of $\chi$ to 
obtain the effective theory in the adiabatic limit. 
Using Eq. \ref{eq:ExpKE} and Eq. \ref{eq:ExpMass},
we obtain,
\be
T_\chi = \alpha\chi^3\rho
\ee 
The energy momentum tensor is again given by Eq. \ref{eq:Tmunu}.
We may also obtain the $\chi$ dependence of the energy density in the 
adiabatic limit. We find, using the procedure described in Ref. \cite{Jain:2011},
\be
\left<{\cal H}\right> \sim {{\bar m_1}\over V}\propto \chi^\alpha
\ee
where $V$ is the volume of space and $\bar m_1 = \chi^\alpha m_1$.
This also implies that
\be
\left<{\partial\phi\over\partial\eta} {\partial\phi\over\partial\eta}\right>
\propto {\chi^\alpha\over \chi^2}
\ee
and hence
\be
\rho\propto {\chi^\alpha\over\chi^4}
\label{eq:rhoNR}
\ee
where we have used Eq. \ref{eq:ExpKE}. Hence we obtain the standard
$1/\chi^3$ evolution for $\alpha=1$. However the energy density decays
faster for $\alpha < 1$.  

Using Eq. \ref{eq:Tmunu} we obtain,
\be
T_{ij} = 0
\ee
and 
\be
T_{00} = \chi^2\left<\partial_0\phi\partial_0\phi\right> = \chi^4\rho
\ee
The gravitational equation of motion then becomes, setting $\bar g_{\mu\nu}
=\eta_{\mu\nu}$,
\be
-2\alpha\chi^{2\alpha-1}{\partial^2\chi\over\partial\eta^2}
-[2\alpha(2\alpha-1)-\xi]\chi^{2\alpha-2}\left({\partial\chi \over \partial\eta}
\right)^2 = {\kappa\over 2}\chi^4\rho
\label{eq:graviNR}
\ee
The equation of motion for $\chi$ becomes,
\be
2\xi\zeta \chi^{2\zeta-1} \left({\partial\chi\over\partial\eta}\right)^2
+ 2\xi\chi^{2\zeta} {\partial^2\chi\over\partial\eta^2} = \kappa \alpha\chi^3\rho
\label{eq:chiNR}
\ee
Eliminating $\rho$ between these two equations, we obtain,
\be
\chi {\partial^2\chi\over\partial\eta^2} = k \left({\partial\chi\over\partial\eta}\right)^2
\ee
where 
\be
k = {\xi -2\alpha^2(2\alpha-1)\over \xi+2\alpha^2}
\ee
This leads to 
\be
{\partial\chi\over \partial\eta} \propto \chi^k
\ee
Using Eq. \ref{eq:chiNR}, we obtain 
\be
\rho\propto \chi^{2\alpha+2k-6}
\ee
Consistency with Eq. \ref{eq:rhoNR} gives,
\be
\xi = 6\alpha^2
\ee
which also gives, $k=1-\alpha/2$.

\subsection{Luminosity Distance}
We next obtain the formula for luminosity distance in this model in order
to make a fit to the high $z$ supernova data.
The main complication in obtaining this formula is the modified redshift
formula, given in Eq. \ref{eq:redshift}. Consider an electromagnetic 
wave emitted by the source at time $t_1$ and observed at time $t_0$. 
Let the luminosity of the source be denoted by $L$.
The 
electromagnetic vector potential is given by
Eq. \ref{eq:EMWave} with $\omega_e$ equal to a constant. Let the wave
propagate radially. Using $ds^2 = dt^2-\chi^2dr^2$ for spatially flat 
Universe, we obtain,
\be
\int_{t_1}^{t_0} {dt\over \chi(t)} = \int_0^{r_1} dr = {\rm const}
\label{eq:r1}
\ee 
for the propagation of light.
Let us consider radiation emitted over a small time interval $dt_1$
and observed over the time interval $dt_0$.
We obtain,
\be
{dt_0\over dt_1} = {\chi(t_0)\over \chi(t_1)}
\ee
The observed flux, $F$, is given by,
\be
F = {Ldt_1\over dt_0} {1\over 4\pi \chi(t_0)^2r_1^2} {\omega(t_0)\over 
\omega(t_1)} 
\label{eq:flux}
\ee
We also have
\be
{\omega(t_0)\over \omega(t_1)} = {\chi(t_1)\over \chi(t_0)}
\ee

So far we have followed the standard text book derivation. 
The main
point of departure arises when we replace the ratio of scale factors
in terms of redshift using Eq. \ref{eq:redshift}. 
We have
\be
{\chi(t_1)\over \chi(t_0)} = {1\over (1+z)^{1/\alpha}}
\ee
Furthermore since the masses scale in our theory, we can no longer 
assume that the peak luminosity of supernovae Type 1a in early Universe is 
same as those of supernovae exploding today.
In the present model we have assumed that all the masses scale in exactly
the same manner. Hence all the dimensional quantities will scale 
according to their mass dimensions. This implies that peak luminosity
scales as,
\be
L \propto \tilde m^2 = \left(\chi^{\alpha-1} m\right)^2
\ee
Let $L_0$ be the peak luminosity of supernova Type 1a exploding at $z=0$. 
We need to insert the factor 
\be
{L_0\over L} = \left({\chi(t_0)\over \chi(t)}\right)^{2\alpha-2}  
\ee
in the formula for the observed 
flux, Eq. \ref{eq:flux}, where $L$ is the peak luminosity of the supernova
exploding in the early Universe.
Using the definition of the luminosity distance, $d_L$, 
\be
F = {L_0 \over 4\pi d_L^2 } 
\label{eq:flux1}
\ee
we obtain,
\be
d_L = (1+z) r_1
\label{eq:dL1}
\ee
where $r_1$ is given by Eq. \ref{eq:r1}.
 The scale parameter in terms
of cosmic time, $t$, is given by
\be
\chi(t) = \left[1-C(t_0-t)\right]^{1/(2-k)}
\ee
where $C$ is a constant of integration.
This gives,
\be
r_1 = {2-k\over C(1-k)}\left[1-{1\over (1+z)^{1/2}}\right]
\ee
where we have used $(1-k)/\alpha = 1/2$.
Substituting this into the formula for $d_L$ and identifying the Hubble 
constant, $H_0$, we obtain,
\be
d_L = (1+z) {2\over H_0} \left[1-{1\over (1+z)^{1/2}}\right]
\ee
Hence we obtain the standard 
formula for non-relativistic matter in Big Bang Model, despite the fact
that our model is very different and even involves time varying masses.
The reason for this is presumably the generalized conformal symmetry imposed
on the model. Since we obtain the same result as the standard Big Bang
cosmology, it is clear that we need some additional component, such
as vacuum energy, to fit the high $z$ supernova data. Alternatively
we need to break conformal invariance. 

\subsection{Supernova stretch factors}
We should emphasize that the supernova light curve stretch factors also
scale in exactly the same manner as in standard Big Bang cosmology, 
despite the fact that our model involves time varying masses. This is
due to the conformal symmetry in our model. We see this explicitly as follows.
Consider, for example, the R-band supernova light curve, given by \cite{Goldhaber},
\be
{I(t)\over I_{max}} = f_R\left((t-t_{max})/w\right) + b
\ee
where $I_{max}$ and $t_{max}$ are respectively the intensity and
time at maximum. A fit to the light curves at different redshifts
yields $w = s(1+z)$, where $s$ is the universal 
stretch factor, independent of $z$. This shows that the observed time scale over
which the supernova intensity decays agrees with the expectation of Big 
Bang cosmology. This need not apply in models where the supernova intensity
intrinsically scales with $z$. In our case we expect that the light
curve would scale as,
\be
{I(t)\over I_{max}} = f_R\left((t-t_{max})\chi/w\right) + b
\ee
Furthermore all dimensional parameters scale with their corresponding
mass dimension. Hence we expect that $w$ which has dimensions of time,
scales as,
\be
w \propto {1\over \tilde m}\propto {1\over \chi^{\alpha-1}}
\ee
Hence we expect,
\be
{I(t)\over I_{max}} = f_R\left((t-t_{max})\chi^\alpha/s\right) + b
\ee
where $s$ is a $z$ independent stretch factor. 
Next using Eq. \ref{eq:redshift}, $\chi^\alpha(t) = 1/(1+z)$,
we see that we obtain the standard dependence on $z$. 

\subsection{Non-relativistic matter and vacuum energy dominated Universe}
We next obtain the formula for luminosity distance including both the
non-relativistic matter and vacuum energy. The action for vacuum energy,
assuming conformal invariance, is given in Eq. \ref{eq:Lambda}. This term
is essentially proportional to $\chi^{4\alpha}$. We point out that
quantum corrections from the matter action will also generate a term
of this type. This is easily seen by considering the effective 
action in Minkowski space, expressed in terms of barred variables, 
such as $\bar m_1 = \chi^\alpha m_1$, 
as discussed in Ref. \cite{Jain:2011}. Since 
all mass parameters are scaled in exactly the same manner, it is clear
that the vacuum energy term, which is proportional to $\bar m_1^4$, 
must scale as $\chi^{4\alpha}$. This implies that, irrespective of
the value of $\alpha$, we shall generate a large cosmological constant
due to quantum corrections, leading to the familiar fine tuning problem.   

The expression for $T_\chi$ gets an additional contribution from the vacuum
energy term. The equation of motion for $\chi$ becomes,
\be
2\xi\zeta \chi^{2\zeta-1} \left({\partial\chi\over\partial\eta}\right)^2
+ 2\xi\chi^{2\zeta} {\partial^2\chi\over\partial\eta^2} = \kappa \alpha\chi^3\rho + 4\kappa\alpha\Lambda\chi^{4\alpha-1}
\label{eq:chiNRVac}
\ee
The gravitational equation of motion remains unchanged. We again find that
for consistent cosmic evolution, $\xi=6\alpha^2$, same as was found in the case of pure non-relativistic matter dominated Universe.
We set 
\be
\rho = \rho_0{\chi^\alpha\over \chi^4}
\ee
The luminosity distance may be obtained by using Eq. \ref{eq:dL1},
where $r_1$ is given by 
\be
r_1 = \int_0^z {dz'\over \alpha H(z')} {1\over (1+z')^{1-1/\alpha}}
\ee
where $H(z)$ is the Hubble parameter. 
We obtain,
\be
d_L = {1+z\over H_0} \int_0^z dz'\sqrt{{1+\Lambda/\rho_0\over
(1+z')^3+\Lambda/\rho_0}}\ .
\ee
Hence we obtain a result independent of $\alpha$. This means that for
all values of $\alpha$ we obtain as good a fit to the high $z$ supernova
data as obtained in the case of standard big bang cosmology. We need
some other cosmological observable, such as CMBR or large scale 
structures to distinquish models with different $\alpha$. We postpone this
for further research. 

\section{Models with broken generalized conformal invariance}
We next consider some models which contain some terms which break 
the generalized conformal 
invariance, discussed in section 4. 
Here we focus primarily on two models, discussed in Ref. \cite{Jain:2011}.
It was shown in Ref. \cite{Jain:2011} that it is possible to fit
the supernova high $z$ data purely in terms of  
generalized cosmological constant or generalized non-relativistic matter. 
We consider the addition of such components in our theory and determine
the fit to supernova data, including also visible matter. 
We shall assume that the visible matter terms as well as the gravitational
terms display generalized conformal invariance with $\alpha=1$. 
These models assume that the standard cosmological constant term 
is identically zero. Hence this partially solves the problem of
fine tuning of the cosmological constant, although we do not have any
reason for this term to be absent. The generalized cosmological
constant terms that are present do not lead to any fine tuning 
since this term is not generated by the matter action at any order 
in the perturbation theory. Furthermore these models solve the 
coincidence problem of dark energy and dark matter since the dominant
cosmological evolution in obtained only from a single component.

\subsection{Generalized cosmological constant}
Here we consider a model based on the generalized cosmological 
constant, discussed in Ref. \cite{Jain:2011}. The basic idea
is to assume a cosmological constant term proportional to $\chi^\delta$
instead of the standard $\chi^4$. Then we may determine $\delta$ by fitting
the high $z$ supernova data. It was found that a consistent cosmological
solution is obtained only if $\delta=\xi-2$. The model provided a good
fit to supernova data only in terms of this single component. The best
fit value of $\xi$ was found to be 4.76.  
We are now interested in determining how the visible matter modifies this
fit. For visible matter, we shall assume that $\alpha=1$. 
Hence the action is now taken as,
\be
 \mathcal{S} = \int d^4x \sqrt{-\bar g}\left[{\chi^{2}\over \kappa} \bar R - 
{\xi\over \kappa}
\bar g^{\mu\nu} \partial_\mu\chi\, \partial_\nu\chi\right] +\mathcal{S}_M 
+\mathcal{S}_\Lambda
\label{eq:actionGS1}
\ee
with $S_M$  
given by Eq. \ref{eq:GS_matter} and Eq. \ref{eq:GS_fermions} with $\alpha=1$.
However the action corresponding to cosmological constant deviates
from that given by Eq. \ref{eq:Lambda}. The generalized cosmological
constant term is taken to be
\be
 \mathcal{S}_\Lambda = -\int d^4x \sqrt{-\bar g}\left[\Lambda \chi^{\xi-2} +
\Lambda_1\chi\right]
\ee
Here the first term is same as in Ref. \cite{Jain:2011}. The second term
is required in order to obtain a consistent cosmological evolution
in the presence of visible matter with $\xi$ different from 6. As seen
in Ref. \cite{Jain:2011}, the unimodular constraint demands that the
power of $\chi$ in the mass term is related to $\xi$. In the present
case we have fixed the power of $\chi$ in the mass term whereas 
$\xi$ can deviate from the value 6. This requires the  
introduction of an additional term proportional to $\Lambda_1$.
We emphasize that none of the terms in $S_\Lambda$ would be generated 
by the matter action at any order in perturbation theory. The matter
can only generate a cosmological constant term proportional to 
$\bar m^4$. Hence it will only generate a term proportional to $\chi^4$.

The equation of motion for $\chi$ is given by,
\be
2\xi{\partial^2\chi\over \partial\eta^2} = \kappa\chi^3\rho
+\kappa\left[\Lambda(\xi-2)\chi^{\xi-3} + \Lambda_1\right]
\ee 
The gravitational equation of motion is same as  
Eq. \ref{eq:graviNR} with $\alpha = 1$.
The energy density of visible matter can be expressed as,
$\rho=\rho_0/\chi^3$. 
 We obtain consistent cosmological
evolution if
\be
\rho_0 = 2\Lambda_1{\xi-3\over 6-\xi}
\ee
This shows that for $\xi=6$, we must set $\Lambda_1=0$. 

The luminosity distance in the present model is given by
$d_L=(1+z) r_1$, where,
\be
r_1 = \int_0^z {dz'\over H(z')}
\label{eq:r1_standard}
\ee 
and the Hubble parameter is given by
\be
H = {H_0\chi^{(\xi/2)-3}\over \sqrt{1+\xi\rho_0/(2\Lambda(\xi-3))}}
\left[ 1+{\xi\rho_0\over 2\Lambda(\xi-3)}\chi^{3-\xi}\right]^{1/2}
\ee
with $\chi=1/(1+z)$. Here $H_0$ is the Hubble constant at present time. 
It is related to the $\Lambda$ and $\rho_0$ by the formula,
\be
H_0 = \sqrt{\kappa\Lambda\over \xi}\left[1+ {\xi\rho_0\over 2\Lambda (\xi-3)}\right]^{1/2}
\ee
The model has three parameters, $H_0$, $\xi$ and $\rho_0/\Lambda$. As
mentioned earlier, setting $\rho_0=0$ provides a good fit to supernova 
data. Here we determine how the fit changes if we include visible matter. 
The visible matter is expected to contribute less than 5\% of the total
energy density of the Universe. 
Hence we fit the data by varying $\rho_0/\Lambda$ from 
0 to 0.05. We find that the best fit values of $\xi$ and $H_0$ vary 
from 4.76 to 4.60 and 69.4 to 74.2 Km/(sec Mpc) respectively as
$\rho_0/\Lambda$ varies from 0 to 0.05. The corresponding $\chi^2$ changes
from 549.2 to 550.09. Hence in the entire range we obtain a good fit 
to the data with $\chi^2$ per degree of freedom less than unity.
Here we have used the data set of 557 Type Ia supernovae
given in Ref. \cite{Amanullah}.

\subsection{Generalized non-relativistic dark matter}
We next consider a model based on generalized non-relativistic dark matter,
discussed in Ref. \cite{Jain:2011}. In this case we assume that the mass
term for dark matter involves a power of $\chi$ different from 4, 
which is expected for the standard covariant model. In Ref. \cite{Jain:2011}
it was shown that consistent cosmological evolution is obtained only
if the power is $\xi-2$. This single component model was also found to
give a good fit to the supernova data, with $\xi = 9.52$. In the present
case we determine how the fit changes if we add visible matter to this model.
As in the previous section, here also we assume that visible matter
action is same as for the covariant model, i.e. $\alpha = 1$ for visible matter.

The action for the present model may be written as,
\be
 \mathcal{S} = \int d^4x \sqrt{-\bar g}\left[{\chi^{2}\over \kappa} \bar R - 
{\xi\over \kappa}
\bar g^{\mu\nu} \partial_\mu\chi\, \partial_\nu\chi\right] +\mathcal{S}_M 
+\mathcal{S}_{DM} + \mathcal{S}_\Lambda
\label{eq:actionGS2}
\ee
Here $S_M$ is
given by Eq. \ref{eq:GS_matter} and Eq. \ref{eq:GS_fermions} with $\alpha=1$.
The dark matter action is given by
\begin{equation}
\mathcal{S}_{DM} = \int d^4x \sqrt{-\bar g}\Bigg[ {\chi^2\over 2}
\bar g^{\mu\nu} (\partial_\mu \psi)(\partial_\nu \psi) - 
{\chi^{\xi-2}\over 2} M^2\psi^2
\Bigg]
\label{eq:GS_darkmatter}
\end{equation}
where $\psi$ refers to the dark matter field. 
The $S_\Lambda$ term is expressed as,
\be
 \mathcal{S}_\Lambda = -\int d^4x \sqrt{-\bar g}\Lambda 
\chi]
\ee
This is not the standard cosmological constant term and is required
in order to obtain a consistent cosmological evolution when visible
matter density is non-zero. As in the previous section, we have set the
standard cosmological constant to be identically equal to zero. Hence this
model also partially solves the cosmological constant problem since 
we do not need to fine tune it to a small value. Furthermore here also 
we obtain the fit dominantly in terms of a single component, i.e.
dark matter. Hence the model also solves the problem of coincidence
of dark matter and dark energy.

The present model treats the visible and dark matter differently. The visible
matter shows no intrinsic evolution with time. In contrast the mass
parameter for dark matter would evolve with time. Hence the redshift
relationship of visible matter and dark matter would differ. For the visible
matter we obtain the usual dependence, 
\begin{equation}
{\omega(t_0)\over \omega'(t_0)} = \left({\chi(t_e)\over \chi(t_0)}\right)
= {1\over 1+z} 
\label{eq:redshift1}
\end{equation}
whereas for dark matter we obtain,
\begin{equation}
{\omega(t_0)\over \omega'(t_0)} = \left({\chi(t_e)\over \chi(t_0)}\right)^{(\xi/2)-2}
\label{eq:redshift2}
\end{equation}
The relationship between the scale factor and redshift is defined in terms
of the visible matter and hence is given by Eq. \ref{eq:redshift1}. Using
this we obtain for dark matter
\be
{\omega(t_0)\over \omega'(t_0)} = \left({1\over 1+z} \right)^{(\xi/2)-2}
\ee
which will deviate from the standard relationship for $\xi\ne 6$. 
This can be tested by observations if the dark matter annihilation
produces a line profile, as expected. Our model predicts that the 
redshift of this line profile should differ from that obtained for visible
matter.

The equation of motion for $\chi$ in the present model can be written 
as,
\be
2\xi {d^2\chi\over d\eta^2} = \kappa {\xi-4\over 2} \chi^3\rho_{DM}
+\kappa\chi^3\rho_M + \kappa \Lambda
\ee
where $\rho_M=\rho_{0}/\chi^3$ and
$\rho_{DM}=\rho'_{0}/\chi^{6-\xi/2}$
 are the energy densities of visible and dark matter respectively.
The gravitational equation of motion is same as 
Eq. \ref{eq:graviNR} with $\alpha = 1$ and $\rho=\rho_{M}+\rho_{DM}$.
We obtain a consistent cosmological solution with
\be
\rho_{0} = 2\Lambda {\xi-3\over 6-\xi}
\ee
The luminosity distance in the present model is given by $d_L=(1+z)r_1$
with $r_1$ same as in Eq. \ref{eq:r1_standard} and the Hubble parameter,
\be
H = {H_0\chi^{(\xi/4)-3}\over \sqrt{1+\xi\rho_{0}/(2\rho'_{0}(\xi-3))}}
\left[ 1+{\xi\rho_{0}\over 2\rho'_{0}(\xi-3)}\chi^{3-\xi/2}\right]^{1/2}
\ee
Here, as in the previous subsection, $\chi=1/(1+z)$. This model also depends
on three parameters, $H_0$, $\xi$, and the ratio of visible to dark
matter, $\rho_0/\rho'_0$. We obtain these by fitting the supernova data. 
The Hubble constant $H_0$ is related to $\rho_0$ and $\rho'_0$ by the formula,
\be
H_0 = \sqrt{\kappa\rho'_0\over \xi}\left[1+ {\xi\rho_0\over 2\rho'_0 (\xi-3)}\right]^{1/2}
\ee
In this case also we expect that $\rho_0/\rho'_0 <0.05$ since 
$\rho_0$ corresponds to visible matter density. We fit to supernova
data \cite{Amanullah} by varying $\rho_0/\rho'_0$ in the range 0 to 0.05. We find that
$\xi$ and $H_0$ vary from 9.52 to 9.34 and 69.4 to 71.9 Km/(sec Mpc) 
respectively as $\rho_0/\rho'_0$ varies in the range 0 to 0.05.
The corresponding $\chi^2$ varies in the range 549.2 to 550.1 which
leads to $\chi^2$ per degree of freedom less than one over the 
entire range.

\section{Spherically Symmetric Solution in Vacuum}
We next determine the spherically symmetric solution in vacuum for the
conformal invariant model corresponding to $\alpha=0$. The main purpose
of this section is determine how this solution may deviate from the
standard Schwarzschild solution. We expect it to show deviation only
at large distances, which can be neglected on solar system scales. 
In the model considered in Ref. \cite{Jain:2011}, we found that solution
is exactly the same as the standard Schwarzschild solution.

Imposing the unimodular constraint
on the metric $\bar g_{\mu\nu}$ we can write it as
\be
\bar g_{\mu\nu} = {\rm diag} \left[{1\over A(r)}, -A(r),-r^2,-r^2\sin^2\theta
\right]
\ee 
The full metric is given by Eq. \ref{eq:gmunu} where $\chi=\chi(r)$. 
The determinant $det[\bar g_{\mu\nu}]$ is equal to the determinant of the Lorentz metric in spherical coordinates. The curvature tensor
satisfies the following equation in vacuum, 
\be
- \left[\bar R_{\mu\nu} - {1\over 4} \bar g_{\mu\nu}\bar R\right]
+\xi\left[\partial_\mu\ln\chi\partial_\nu\ln\chi -{1\over 4}\bar g_{\mu\nu}
\partial^\lambda\ln\chi\partial_\lambda\ln\chi \right] = 0
\label{eq:Rmunu2}
\ee
We have 
\be
{\bar R_{rr}\over A} + A \bar R_{tt} = 0
\ee
which gives,
\be
\partial_r \ln\chi = 0
\ee
Hence $\chi$ is a constant.
This implies that
\be
\bar R_{\mu\nu} - {1\over 4} \bar g_{\mu\nu} \bar R = 0
\ee
This is as far as we can go. The Schwarzschild solution indeed solves 
this equation. However it is not unique. We cannot
set $\bar R=0$. We find,
\be
\bar R'_{\theta\theta} - {2\over r} \bar R_{\theta\theta} = 0
\ee
where the prime refers to derivative with respect to $r$. 
Hence 
\be
\bar R_{\theta\theta} = C_2 r^2
\ee
where $C_2$ is a constant. This implies
\be
rB' + B = 1+C_2r^2
\ee
where $B=1/A$.
If we set $C_2=0$ we get the standard Schwarzschild solution. This also applies
approximately for small $r$. However for large $r$ the solution gets modified.
We find,
\be
B(r) = {1\over A(r)} = 1 + {C_3\over r} + {r^2\over 3}C_2
\ee
where $C_3$ is another constant. By using the relationship $B=1+2\phi$,
where $\phi$ is the gravitational potential, we can relate $C_3$ to the
mass $M$ of the source in the usual manner. We have, neglecting the term 
proportional to $C_2$ for small $r$,
\be
C_3 = -2 GM
\ee
where $G$ is the gravitational constant.
Hence we find the gravitational potential,
\be
\phi = -{GM\over r} + {C_2\over 6} r^2
\ee
We find that the potential deviates from the standard Newtonion potential
at large distances. It is clearly of interest to see if this can explain
the galactic rotation curves. By relating the gravitational force due to this
potential on a test mass in circular motion with speed $v$ at distance $r$ 
from the source, we find that, at large distances, 
\be
v \approx \sqrt{{C_2\over 3}}\ r
\ee
Hence the rotational speed increases linearly with $r$. 
We need to extend this solution to other models discussed in this paper. 
We postpone such a solution to future research. Here we only comment
that the solution is expected to deviate from the standard Schwarzschild
solution only at large distances
due to the presence of the field $\chi$. This might have implications
for galactic rotation curves. 
A fit to the rotation curves 
in an alternate unimodular model is given in Ref.
\cite{SinghNK}

\section{Discussion and Conclusions}
We have considered a class of models which obey only unimodular 
general coordinate invariance. We first considered a model which displays
global conformal invariance. This model, however, fails cosmologically since it 
predicts null redshift. We then introduced a class of models which
display a global symmetry which we termed generalized conformal invariance.
The formula for luminosity distance in these models, in the presence
of dark matter and vacuum energy, turns out to be identical to  
obtained in the standard Big Bang model. This is despite the fact that
cosmic evolution in 
these models is very different and even involves effectively time
varying masses. Hence the fit to supernova
data is not able to distinguish this class of models from the standard
$\Lambda CDM$. We require a fit to some other cosmological 
observable, not pursued in the present model.
We have also shown that, despite the fact that our models involve time
varying masses, they do not disagree with the observed scaling of the 
supernova stretch factors \cite{Goldhaber}. This is because all mass
parameters scale by the scale factor.  

We also   
study two models, which do not display generalized conformal invariance.
These are based on the models presented in Ref. \cite{Jain:2011} where
we showed that a fit to supernova data is possible including only a 
single component, either dark matter or dark energy. In the
case of dark matter we essentially
generalize the mass term corresponding to dark matter so that it can
give a good fit to supernova data. In this model we find that the redshift
dependence of dark matter is different from visible matter. This implies
that line profiles corresponding to dark matter annihilation may show 
different redshift dependence in comparison to spectral lines of visible
matter. This prediction may be tested in future if dark matter line
profiles are observed. We also study another model in which the supernova data
can be fit entirely in terms of a single component corresponding to
a generalized cosmological constant term \cite{Jain:2011}. We extend
the fits in both these models by also including the contribution due
to visible matter. These models are interesting since they do not require
the standard vacuum energy term, thus alleviating the fine tuning problem 
of the cosmological constant. Furthermore since the fit is obtained
dominantly in terms of a single component, these models also solve
the coincidence problem of dark matter and dark energy.

\bigskip
\noindent
{\bf Acknowledgements} \\
We thank Subhadip Mitra, Sukanta Panda, V. Ravishankar and Kandaswamy Subramanian for useful discussions.
PJ acknowledges a useful communication with David Branch.
GK thanks CSIR for financial assistance. 

\begin{spacing}{1}
\begin{small}

\end{small}
\end{spacing}
\end{document}